\documentclass[pre,aps,preprint,superscriptaddress]{revtex4-1}

\usepackage{graphicx}
\usepackage{amsmath}
\usepackage{dcolumn}
\usepackage{txfonts}
\usepackage{bm}
\usepackage{multirow}
\usepackage{booktabs}
\usepackage{epstopdf}

\begin{document}
	\title{Reconfigurable, zero-energy, and wide-temperature loss-assisted thermal nonreciprocal metamaterials}
	
	\author{Min Lei}
	\author{Peng Jin}
	\author{Yuhong Zhou}
	\affiliation{Department of Physics, State Key Laboratory of Surface Physics, and Key Laboratory of Micro and Nano Photonic Structures, Ministry of Education, Fudan University, Shanghai 200438, China}
	
	\author{Ying Li}
	\affiliation{State Key Laboratory of Extreme Photonics and Instrumentation, Key Laboratory of Advanced Micro/Nano Electronic Devices $\&$ Smart Systems of Zhejiang, Zhejiang University, Hangzhou 310027, China}
	\affiliation{International Joint Innovation Center, The Electromagnetics Academy of Zhejiang University, Zhejiang University, Haining 314400, China}
	\affiliation{Shaoxing Institute of Zhejiang University, Zhejiang University, Shaoxing 312000, China}
	
	\author{Liujun Xu}\email{ljxu@gscaep.ac.cn}
	\affiliation{Graduate School of China Academy of Engineering Physics, Beijing 100193, China}
	
	\author{Jiping Huang}\email{jphuang@fudan.edu.cn}
	\affiliation{Department of Physics, State Key Laboratory of Surface Physics, and Key Laboratory of Micro and Nano Photonic Structures, Ministry of Education, Fudan University, Shanghai 200438, China}
	
	\begin{abstract}
		Thermal nonreciprocity plays a vital role in chip heat dissipation, energy-saving design, and high-temperature hyperthermia, typically realized through the use of advanced metamaterials with nonlinear, advective, spatiotemporal, or gradient properties. However, challenges such as fixed structural designs with limited adjustability, high energy consumption, and a narrow operational temperature range remain prevalent.
		Here, a systematic framework is introduced to achieve reconfigurable, zero-energy, and wide-temperature thermal nonreciprocity by transforming wasteful heat loss into a valuable regulatory tool. Vertical slabs composed of natural bulk materials enable asymmetric heat loss through natural convection, disrupting the inversion symmetry of thermal conduction.
		The reconfigurability of this system stems from the ability to modify heat loss by adjusting thermal conductivity, size, placement, and quantity of the slabs. Moreover, this structure allows for precise control of zero-energy thermal nonreciprocity across a broad temperature spectrum, utilizing solely environmental temperature gradients without additional energy consumption.
		This research presents a different approach to achieving nonreciprocity, broadening the potential for nonreciprocal devices such as thermal diodes and topological edge states, and inspiring further exploration of nonreciprocity in other loss-based systems.
	\end{abstract}
	
	\maketitle
	\section*{Introduction}
	\begin{figure}[t]
		\includegraphics[width=.9\linewidth]{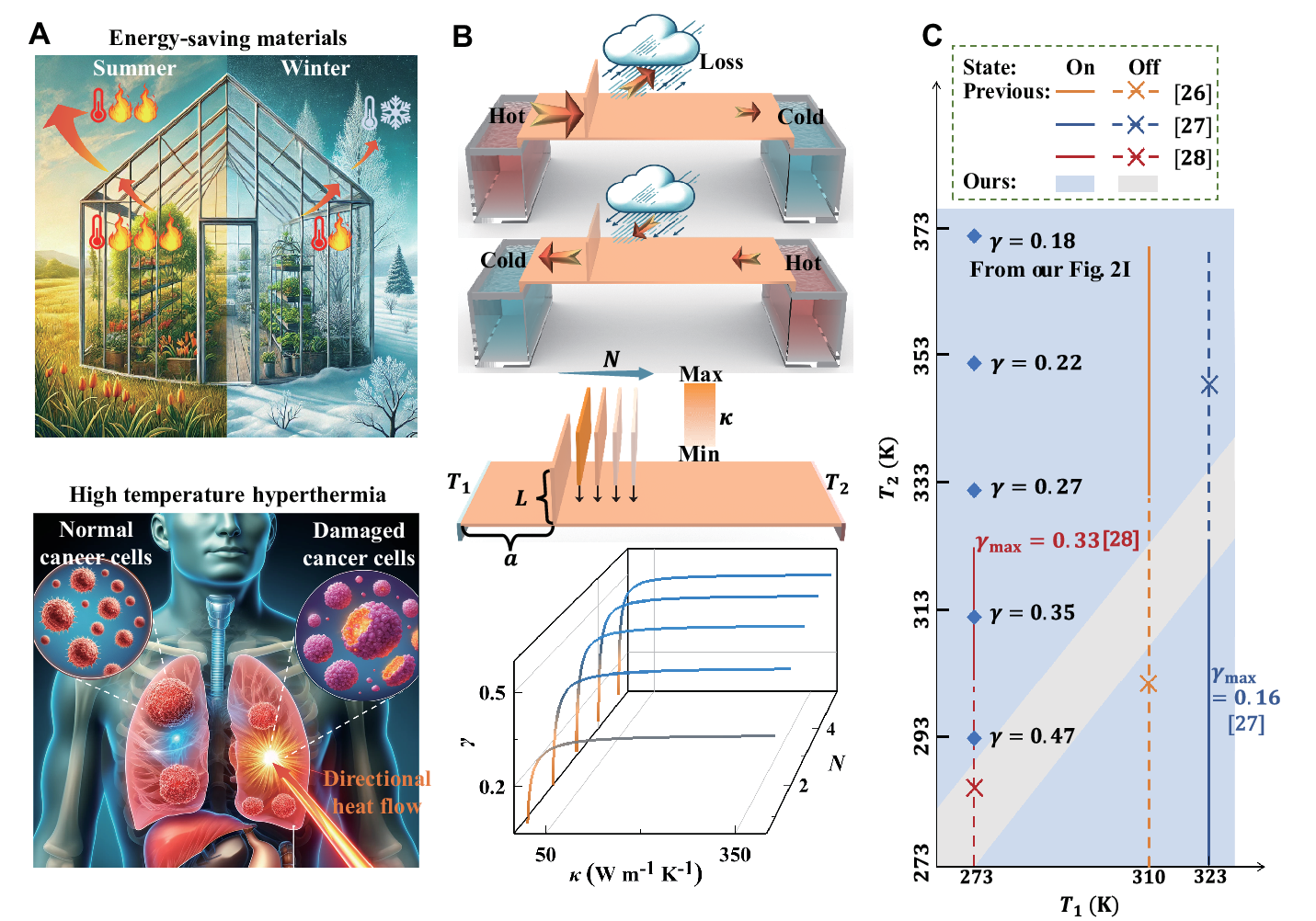}\\
		\caption{Principle of loss-assisted thermal nonreciprocal metamaterials. 
			({\it A}) Illustrations of common usage scenarios related to general thermal nonreciprocity. Upper panel: The adjustable nonreciprocal thermal design enables efficient thermal management of enclosed spaces throughout all seasons, reducing reliance on high-energy consumption devices like air conditioning. Take a greenhouse as an example: the temperature difference caused by environmental fluctuations can be minimized through adjustable structural design. In summer, it enhances heat dissipation, while in winter, it reduces heat loss, dynamically maintaining the optimal indoor temperature for vegetable greenhouses. Lower panel: The thermal rectification achieved by controlling the direction and distribution of heat can facilitate precise thermotherapy for treating cancer. These two potential applications are described in more detail in Fig. S11.
			({\it B}) Schematic of the loss-assisted nonreciprocal metamaterial. In asymmetric structures made from natural bulk materials, natural convection-induced asymmetric thermal losses disrupt the inherent spatial symmetry of thermal conduction. The rectification ratio $\gamma$ can be adjusted based on parameters such as thermal conductivity $\kappa$, quantity $N$, height $L$, and position $a$ of vertical plates.
			({\it C}) Comparison of operable temperature ranges between previous studies and this nonreciprocal work, where 'on' and 'off' indicate the possibility and impossibility of achieving effective thermal rectification, respectively. Unlike earlier research on nonlinear nonreciprocal implementations, our nonreciprocal devices are not limited to specific temperature points. As long as the temperature difference exceeds 10 K ($\delta T=|T_1-T_2|\geq 10$ K), thermal rectification output can be effectively generated within the natural ambient temperature range of 273 K to 373 K. The $\gamma$ data in ({\it B}) and ({\it C}) are derived from the outflow rectification ratio $\gamma_2$ in Fig. \ref{FigLM2}.}
		\label{FigLM1}
	\end{figure}
	nonreciprocity refers to the differing responses observed when processes occur in forward versus reverse directions. Thermal nonreciprocity is vital for applications such as chip cooling \cite{ErpNat20}, waste heat recovery \cite{YanNM22}, high-temperature hyperthermia treatments \cite{ZhuCR23}, and energy-efficient material design \cite{ShaoNS24}. The asymmetric heat transfer achieved by thermal nonreciprocity can control the temperature of enclosed spaces and treat diseases through directional heat dissipation and directional heating respectively (Fig. \ref{FigLM1}\emph{A}). Through the use of metamaterials \cite{FanAPL08,ZhangNRP23,YangRMP24,XuNCS24,JinResearch23,ChenINN22,ChenPNAS22,ChenDro23,PendryAP22,PendryNC21,BosPNAS21,HavPNAS24,PendryPNAS20,PendryPRL19}, various functions of thermal nonreciprocity have been realized, including thermal diodes \cite{LiRMP12}, transistors \cite{BenPRL14}, logic circuits \cite{WangPRL07}, and topological edge states \cite{XuEPL21,YoSR21}. Achieving thermal nonreciprocity often involves metamaterials with nonlinear \cite{LiPRL15,WangPCCP17,PalMH15,ZhouNC23,JinAM24,XuPNAS23}, advective \cite{JuAM23,JuAM24,XuPRL23,LiCPL21,LiAM20,XuAPL21,XuPRL22-1,JinPNAS23}, spatiotemporal \cite{TorPRL18,LiNC22,XuPRL22-2,LeiMTP23,PendryPNAS19,PendryPRL19-2}, or gradient properties \cite{XuNSR23,CaoCP21,CaoCPL22,LiuSB,SuPRAP23}. Despite these advancements, three major challenges hinder broader application. First, gradient materials can produce stable rectified outputs through simulated advection \cite{XuNSR23}, yet their rigid structures lack the flexibility required for diverse practical applications. Second, while metamaterials influenced by convection or spatiotemporal modulation can effectively manage nonreciprocal heat transfer under external forces \cite{JuAM24,XuPRL22-2}, they typically require additional energy, making the pursuit of energy-free solutions highly desirable. Lastly, nonlinear materials, especially shape memory alloys, are excellent at maximizing rectification effects in unidirectional heat conduction \cite{LiPRL15}, but their effectiveness is limited to temperatures near their phase transition points, restricting their usable temperature range. Addressing these challenges is crucial in advancing metamaterials science and thermal engineering to develop nonreciprocal thermal designs that are reconfigurable, operate without energy consumption, and function across a broad temperature range.

	Studies in optics and acoustics have inspired new approaches for thermal nonreciprocal design \cite{CalPRAP18,KadNRP19,FleSCI14,GuoLSA19,PendryLSA22}. A novel method was recently introduced in optical contexts to leverage energy loss—typically considered detrimental due to its impact on energy efficiency, equipment lifespan, and control complexity—for achieving optical nonreciprocity \cite{HuangLSA21}. Similarly, thermal nonreciprocity has been innovatively achieved by introducing asymmetric thermal radiation loss in gradient materials \cite{SuPRAP23}. Transforming adverse thermal losses into a potent tool can effectively enhance energy efficiency through thermal nonreciprocity. However, due to the non-tunable nature of gradient materials and the substantial need for thermal radiation, the tunability and range of nonreciprocal modulation remain constrained. Thus, there is significant value in exploring adjustable structural designs based primarily on thermal losses.

	Here, a framework has been developed that utilizes heat loss for reconfigurable, zero-energy thermal nonreciprocity across a broad temperature range. This framework employs asymmetrically distributed vertical slabs (Fig. \ref{FigLM1}\emph{B}). The inherent asymmetry in heat loss \cite{LiPRB21}, naturally occurring between the slabs and the ambient environment, disrupts the inherent symmetry of thermal conduction. Natural bulk materials offer flexible, reconfigurable control over various parameters, including thermal conductivity, size, position, and the number of slabs. The rectification ratio, which has a maximum value of 1, quantifies the nonreciprocal effect and has been verified through simulations and experiments. The rectification ratio of the stainless steel sample can be freely adjusted between 0.0 and 0.5. 
	This system doesn't require external energy input; it solely utilizes ambient temperature differences to achieve zero-energy thermal nonreciprocity. 
	This structure is not limited to any specific temperature and can provide effective and adjustable nonreciprocal output over a wide temperature range. Specifically, a rectification ratio greater than 0.1 can be considered effective nonreciprocal output. An effective rectification ratio can be achieved for stainless steel samples when the temperature difference between hot and cold sources exceeds 10 K (Fig. \ref{FigLM1}\emph{C}).
	These findings overcome the limitations of non-adjustable thermal nonreciprocity, providing a new paradigm for leveraging heat loss across a wide temperature range.
	
	\section*{Results}
	\subsection*{Loss-assisted thermal nonreciprocity}
	We perform a detailed analysis of the temperature distribution throughout the entire structure to quantify the nonreciprocal effects and predict variations based on specific parameters. The structure is divided into four distinct sections: three on the base plate (designated as Areas I, II, and III) and one on the vertical plate (Area IV) (Fig. \ref{FigLM2}\emph{A}). Area II acts as a staggered zone between the base plate and the vertical plate. To simulate conditions, heat and cold sources are applied at the extremities of the base plate, assuming linear heat conduction. Additionally, there is a thermal exchange with the surroundings on the vertical plate, aside from the heat transfer originating from the base plate. The vertical plate, serving as an extension to the base plate, exhibits a temperature distribution characterized by an exponential function (\emph{SI Appendix}, Section 1). Considering the vertical plate's negligible thickness relative to the base plate's length, the interface temperature between Regions II and IV is assumed to be uniform. The analysis primarily focuses on the temperature distribution along the $z$-axis in Region IV. The general solution for the thermal field distribution across the four regions is provided by the following equations: 
	\begin{equation}\label{LM1}
		\begin{split}
			T_{\rm I}(x)&=c_1x+c_2,\\
			T_{\rm II}(x)&=c_3x+c_4,\\
			T_{\rm III}(x)&=c_5x+c_6,\\
			T_{\rm IV}(z)-T_{\rm amb}&=d_1e^{mz}+d_2e^{-mz},
		\end{split}
	\end{equation}
	where $c_1, c_2, c_3, c_4, c_5, c_6, d_1, d_2$ are constants determined by ensuring the continuity of temperature and heat flux at the boundaries. The parameter $m=\sqrt{hP/(\kappa A_{\rm c})}$ involves the natural convection coefficient $h$, the thermal conductivity $\kappa$, and the perimeter $P$ of the vertical plate's top edge. The boundary conditions on the contact surface of the four regions can be written as: $
	c_2=T_1,
	c_1a+c_2=c_3a+c_4,
	c_3b+c_4=c_5b+c_6,
	c_5c+c_6=T_2,
	c_1a+c_2-T_{\rm amb}=d_1+d_2,
	-\kappa_{\rm b} Ac_3=-\kappa_{\rm b} Ac_5,
	-\kappa_{\rm b} Ac_1=-\kappa_{\rm b} Ac_3-\kappa A_{\rm c} m(d_1-d_2),
	hA_{\rm c}(d_1e^{mL}+d_2e^{-mL})=-\kappa A_{\rm c} m(d_1e^{mL}-d_2e^{-mL}).$
	Here the positions of the contact surfaces between Area I and II, and Area II and III are denoted as $a$ and $b$, respectively. The base plate's length is $c$, with a cross-sectional area $A$ in the $x$-axis direction. The vertical plate has a cross-sectional area $A_{\rm c}$ in the $z$-axis direction and a height $L$, as illustrated in \emph{SI Appendix}, Fig. S1. The thermal conductivity coefficients of the vertical plate and the base plate are $\kappa$ and $\kappa_{\rm b}$ respectively. The cold and heat source temperatures are $T_1$ and $T_2$ respectively, and the ambient temperature is expressed as $T_{\rm amb}$. 
	The boundary conditions detail the temperature relationships at various contact surfaces, indicating different heat flow paths in the structure.
	Specifically, the fifth boundary condition asserts that the temperature at the contact surface between Region II and IV ($z=0$) equals that between Region I and II ($x=a$). The seventh boundary condition describes the division of heat flow from Area I into Areas II and IV, while the eighth equation presents the boundary condition at the contact surface between Area IV's top edge ($z=L$) and the air. Solving these eight equations can yield the temperature distribution of the entire structure.	
	
	With fixed geometric parameters and material properties, the temperature distribution along the base and the vertical plates is graphically represented (Fig. \ref{FigLM2}\emph{B}-\emph{C}). Both the theoretical and simulation studies use the same parameters, namely $\kappa=\kappa_{\rm b}=16.3~\rm W~m^{-1}~K^{-1}$ (stainless steel), $h=20~\rm W~m^{-2}~K^{-1}$, $T_1=373~\rm K$ and $T_2=273~\rm K$, $T_{\rm amb}=298~\rm K$, $L=0.04~\rm m$, $a=0.01~\rm m$, $b=0.012~\rm m$, $c=0.1~\rm m$, and the structure's width is 0.04 m. This figure uncovers a distinct difference in the temperature profiles in the forward and backward directions along the base plate, thus demonstrating nonreciprocal heat transport. 
	Moreover, the vertical plate shows a consistently higher temperature in the forward direction than in the backward direction. The strength of natural convection correlates directly with the temperature difference between the plate and the surrounding air. This variance in natural convection between opposing directions results in nonreciprocal behavior. The agreement between the simulated temperature distribution and theoretical predictions confirms the accuracy of the theoretical framework.
	
	To more intuitively showcase the nonreciprocal effect, we examine the distribution of heat flow in the base plate along the $x$-axis (Fig. \ref{FigLM2}\emph{B}). With the inclusion of natural convection, the heat flow values in Areas I and III exhibit significant variances between the forward and backward directions. Diverging from prior research where heat flow is generally seen as continuous, our findings indicate sharp transitions in heat flow, maintaining constant levels in Areas I and III. This distinction prompts a separate calculation of inflow and outflow heat values. We introduce the rectification ratios $\gamma_1$ and $\gamma_2$, defined as
	\begin{equation}\label{LM3}
		\begin{array}{c}
			\gamma_1=\frac{|q_{\rm FI}-q_{\rm BIII}|}{|q_{\rm FI}+q_{\rm BIII}|},\gamma_2=\frac{|q_{\rm FIII}-q_{\rm BI}|}{|q_{\rm FIII}+q_{\rm BI}|}.
		\end{array}
	\end{equation}
	Here, $\gamma_1$ represents the discrepancy in heat flow inflow between forward and backward directions, dependent solely on the forward inflow ($q_{\rm FI}=|-\kappa_{\rm b} A \nabla T_{\rm FI}(x)|$) and backward inflow ($q_{\rm BIII}=|-\kappa_{\rm b} A \nabla T_{\rm BIII}(x)|$). The subscripts `F' and `B' denote `Forward' and `Backward'. Similarly, $\gamma_2$ quantifies the difference in outflow heat flow values, involving the forward outflow ($q_{\rm FIII}=|-\kappa_{\rm b} A \nabla T_{\rm FIII}(x)|$) and backward outflow ($q_{\rm BI}=|-\kappa_{\rm b} A \nabla T_{\rm BI}(x)|$). A maximum value of 1 for either ratio signifies a scenario where heat flow is unidirectional, occurring exclusively in the forward direction with no corresponding flow in the backward direction.
	
	\subsection*{Multi-parameter control of thermal nonreciprocity}
	\begin{figure}[t]
		\includegraphics[width=\linewidth]{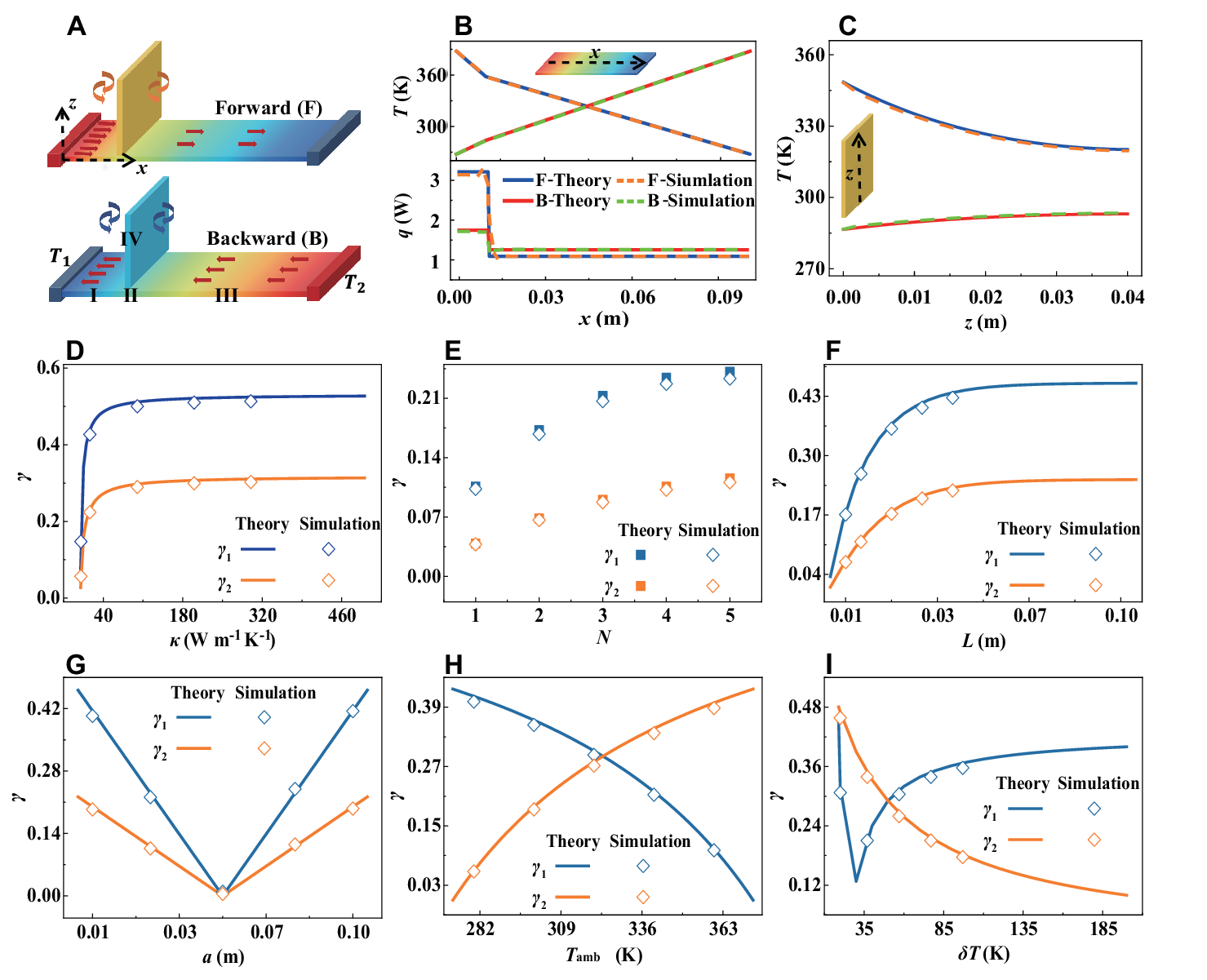}\\
		\caption{Multi-parameter control characteristics of loss-assisted thermal nonreciprocity. ({\it A}) Structural diagram. The entire structure is divided into four areas. ({\it B}) Temperature distribution at the centerline of the base plate and total heat flow distribution in the base plate. ({\it C}) Temperature distribution along the centerlines of the vertical plate. ({\it D})-({\it I}) Rectification ratio ($\gamma_1$ and $\gamma_2$, representing inflow and outflow) changes with thermal conductivity $\kappa$, number $N$, height $L$, and position $a$ of the vertical plates, ambient temperature $T_{\rm amb}$, and the temperature difference $\delta T=|T_1-T_2|$.}
		\label{FigLM2}
	\end{figure}
	
	Our research demonstrates that an asymmetric structural design, combined with natural convection, enables nonreciprocal heat transport. This effect is influenced by several factors, including the thermal conductivity ($\kappa$) of the vertical plate. We adjust $\kappa$ while keeping other parameters shown in Fig. \ref{FigLM2}\emph{B} constant. The rectification ratio varies with the thermal conductivity $\kappa$ of the vertical plate (Fig. \ref{FigLM2}\emph{D}). As $\kappa$ increases, both the inflow and outflow rectification ratios increase, approaching their maximum value. This progression stems from the direct relationship between thermal conductivity and the temperature difference across the vertical plate during heat transport phases; a higher $\kappa$ results in an increased temperature of the vertical plate during forward transport and a decreased temperature during backward transport. Under constant environmental temperature, the greater the temperature difference between the forward and backward directions of the vertical plate, the more pronounced the nonreciprocal effect achieved by natural convection.
	Through simulations employing thermal conductivities of 0.5, 16.3, 100, 200, and 300 $\rm W~m^{-1}~K^{-1}$, we analyze the heat flow values for inflow and outflow (with detailed results in \emph{SI Appendix}, Fig. S2). The rectification ratios derived from these simulations concur with our theoretical predictions (Fig. \ref{FigLM2}\emph{D}), thus corroborating our theoretical framework. Fig. \ref{FigLM2}\emph{D} also reveals that when the thermal conductivity of the vertical plate significantly surpasses that of the base plate, the rectification ratio's adjustable scope is constrained by the base plate's thermal conductivity. Consequently, in applications where the structure is uniformly composed of a single material, modifying the structure's overall thermal conductivity markedly broadens the rectification ratio's adjustable range. Contrary to the rectification ratio's direct correlation with the vertical plate's thermal conductivity, it inversely correlates with the overall structure's thermal conductivity (\emph{SI Appendix}, Section 3). For instance, at a thermal conductivity ($\kappa$) of 0.1 $\rm W~m^{-1}~K^{-1}$, the rectification ratios ($\gamma_1$ and $\gamma_2$) are recorded at 0.76 and 0.68, respectively. In contrast, at $\kappa=400~\rm W~m^{-1}~K^{-1}$, the ratios drop to 0.06 and 0.02, demonstrating that significant modifications to the rectification ratio can be achieved by adjusting the overall thermal conductivity.
	
	The rectification ratio can be precisely adjusted by altering the number ($N$) of vertical plates, according to our theoretical framework. Each plate is uniformly considered in the model, and adding a vertical plate introduces six new variables due to the creation of three additional regions in the structure. We calculate the temperature distribution across the entire structure by resolving these simultaneous boundary conditions, with detailed explanations provided in \emph{SI Appendix}, Section 4. Consistent parameters in our simulations allow us to visualize the heat flow throughout the structure and determine the rectification ratio, with results showcased in \emph{SI Appendix}, Fig. S4.
	Fig. \ref{FigLM2}\emph{E} demonstrates how the number of vertical plates affects the rectification ratios when the structure has homogeneous material properties and a thermal conductivity of $\kappa=200 \rm ~W~m^{-1}~K^{-1}$. The theoretical and simulated outcomes closely match, showcasing an average numerical discrepancy of only 3\%, which underscores the high accuracy of our model. It is observed that the rectification ratios for both inflow and outflow increase with the addition of plates, indicating that each new plate amplifies the disparity in natural convection during forward and backward heat transport. However, this relationship changes with variations in thermal conductivity. For instance, at a thermal conductivity of $\kappa=2.1 \rm ~W~m^{-1}~K^{-1}$, the rectification ratio initially rises with the addition of plates, peaks with two plates, and then diminishes, as detailed in \emph{SI Appendix}, Fig. S5.
	Under general conditions such as Fig. \ref{FigLM2}\emph{C}, during forward heat transfer, the temperature of vertical plates is generally higher than the ambient temperature, facilitating outward natural convection. Conversely, during backward transfer, they are cooler than the environment, resulting in inward natural convection. However, in situations with lower thermal conductivity and an increased number of vertical plates, the temperature on the vertical plates near the center during forward heat transfer can fall below that of the surrounding environment. This leads to a reduction in outward natural convection from all vertical plates, subsequently decreasing the rectification ratio. Therefore, the relationship between the rectification ratio and the number of plates is influenced by their thermal conductivity, underscoring the necessity of balancing the number of plates and their thermal conductivity to achieve the optimum rectification ratio.
	
	Several factors influence the rectification ratio, including the height of the vertical plate ($L$), its position ($a$), ambient temperature ($T_{\rm amb}$), and the temperature differential ($\delta T$) between the cold and hot sources. We focus solely on the impact of individual parameters on the rectification ratio while holding all other factors constant. We maintain a constant thermal conductivity of $\kappa=16.3~\rm W~m^{-1}~K^{-1}$ and use a single vertical plate ($N=1$). The effects on the rectification ratio are documented in Fig. \ref{FigLM2} \emph{F}-\emph{I}, with comprehensive simulation details in \emph{SI Appendix}, Section 6-9. Increasing $L$ expands the convection area, thus enhancing the rectification ratio by accentuating convection differences between forward and backward directions. Nevertheless, this enhancement reaches a limit at a certain height because heat transport along the vertical plate eventually stabilizes with the ambient temperature. Further increasing the plate's length has minimal impact on the rectification ratio. Conversely, extending the vertical plates across the base plate can amplify natural convection and thereby increase the rectification ratio (\emph{SI Appendix}, Fig. S6 G-I).
	The position of the vertical plate is also pivotal in determining the rectification ratio. Plates closer to the center yield lower ratios due to reduced asymmetry, with centrally placed plates achieving symmetrical, reciprocal heat transport, resulting in theoretical and practical rectification ratios of zero. The ambient temperature ($T_{\rm amb}$) inversely affects the inflow and outflow rectification ratios—decreasing the former and increasing the latter. When $T_{\rm amb}$ matches the mean temperature of the cold and hot sources, the rectification ratios ($\gamma_1$ and $\gamma_2$) equalize.
	Moreover, holding the temperature of one source ($T_1$) constant while raising the other ($T_2$), thus increasing $\delta T=|T_1-T_2|$, decreases the outflow rectification ratio. The inflow rectification ratio initially drops and then rises with $\delta T$. If the overall system temperature is below $T_{\rm amb}$, air warms the vertical plate, reducing the inflow rectification ratio as $\delta T$ increases and diminishes the heating effect. Conversely, if the system temperature exceeds $T_{\rm amb}$, the air cools the plate, and as $\delta T$ increases, the cooling effect intensifies, boosting the inflow rectification ratio. While natural convection can either be loss or gain, it typically results in a loss in real-world scenarios. Unlike shape memory alloys, which are limited to specific temperature ranges for nonreciprocity~\cite{LiPRL15}, our design for thermal nonreciprocity is widely adaptable across a broad temperature spectrum and can be easily adjusted. A rectification ratio greater than 0.1 signifies substantial nonreciprocal temperature distribution, defined as effective thermal nonreciprocity. As demonstrated in Fig. \ref{FigLM2}\emph{H}, with a fixed temperature difference of 100 K between the heat sources, both the inflow and outflow rectification ratios exceed 0.1 across environmental temperatures from 286 K to 360 K, confirming effective nonreciprocal output. At lower or higher ambient temperatures, the nonreciprocity of heat flow into or out of the system becomes more pronounced. This suggests that our nonreciprocal design is universally applicable in natural settings. With the environmental temperature set at 298 K, and temperature differences ranging from 20 K to 200 K, we consistently achieve effective rectification ratios (Fig. \ref{FigLM2}\emph{I}). 
	In practical applications, both the temperature difference between hot and cold sources and natural convection can be realized by the ambient temperature difference, eliminating the need for external power mechanisms. This approach achieves zero-energy nonreciprocity in heat transfer, markedly reducing energy consumption and boosting efficiency. 
	
	\subsection*{Numerical and experimental verification}
	\begin{figure}[t]
		\includegraphics[width=.9\linewidth]{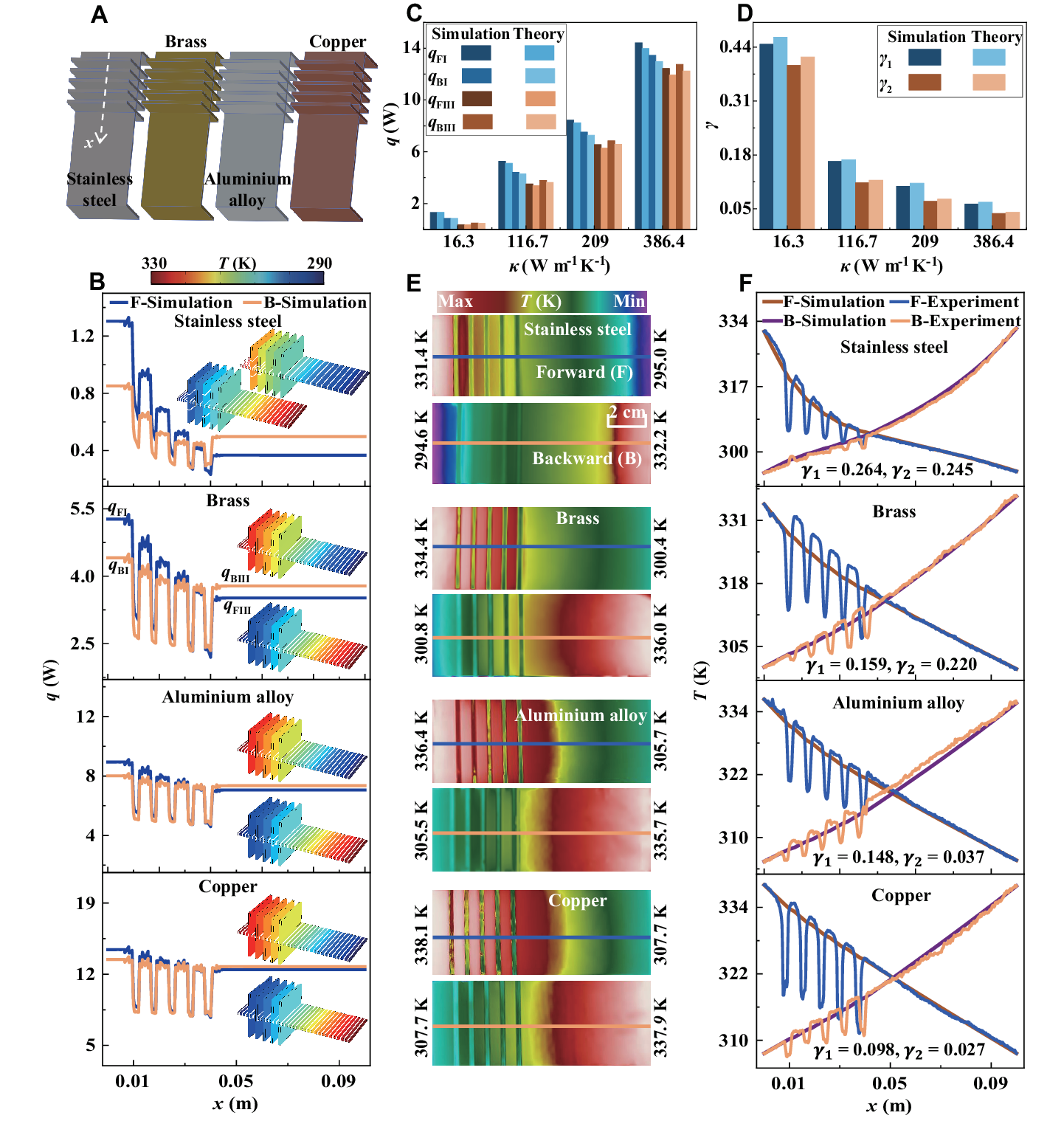}\\
		\caption{Theoretical, simulated, and experimental results of nonreciprocal heat transport configurations with different thermal conductivity. ({\it A}) Structural diagrams of four material configurations. ({\it B}) Simulation outcomes for four different materials: stainless steel ($\kappa=16.3 ~\rm W~m^{-1}~K^{-1}$), brass ($\kappa=116.7~\rm W~m^{-1}~K^{-1}$), aluminum alloy ($\kappa=209 ~\rm W~m^{-1}~K^{-1}$), and copper ($\kappa=386.4 ~\rm W~m^{-1}~K^{-1}$). ({\it C}) Analysis of how heat flow values change with thermal conductivity, covering forward inflow ($q_{\rm FI}$), forward outflow ($q_{\rm FIII}$), back inflow ($q_{\rm BIII}$), and back outflow ($q_{\rm BI}$). These values are derived from the simulations shown in Fig. \ref{FigLM3}\emph{B}. ({\it D}) The variation of inflow rectification ratio ($\gamma_1$) and outflow rectification ratio ($\gamma_2$) with thermal conductivity. ({\it E}) Experimental temperature distribution of the four materials. ({\it F}) A comparative analysis of the temperature distribution along the base plate centerline between experimental results and simulation data. The parameters employed in the simulations are consistent with those used in the experiments.}
		\label{FigLM3}
	\end{figure}
	
	Through simulations and experiments, we validate the reconfigurable nonreciprocity of this asymmetric structure in natural convection settings. Our study emphasizes the critical factors influencing the rectification ratio: thermal conductivity and the quantity of vertical plates. By designing vertical plates to extend through the base plates, we optimize natural convection across vertical panels and improve the structure's stability. Our findings reveal that increasing the number of vertical plates in materials with high thermal conductivity significantly boosts the rectification ratio. We opt for a design featuring five vertical boards that intersect the baseboard. Moreover, constructing each sample from a uniform material minimizes heat loss from thermal contact between different materials while achieving a high rectification ratio.
	
	We examine samples constructed from four distinct materials (Fig. \ref{FigLM3}\emph{A}): stainless steel ($\kappa=16.3 ~\rm W~m^{-1}~K^{-1}$), brass ($\kappa=116.7~\rm W~m^{-1}~K^{-1}$), aluminum alloy ($\kappa=209 ~\rm W~m^{-1}~K^{-1}$), and copper ($\kappa=386.4 ~\rm W~m^{-1}~K^{-1}$). In the simulations, all base plate surfaces, aside from the $x$-directional cold and heat sources, are assumed thermally insulated. The vertical plates experience natural convection at an ambient temperature of $T_{\rm amb}=305$ K and a convection coefficient of $h=8 \rm ~W~m^{-2}~K^{-1}$. We set the heat and cold source temperatures to 330 K and 290 K, respectively. Fig. \ref{FigLM3}\emph{B} shows the resulting simulated temperature distributions and thermal streamline diagrams, where both forward and backward heat flows in all samples display pronounced nonreciprocal characteristics.
	
	Analyzing heat flow in Regions I and III, as compared to theoretical expectations depicted in Fig. \ref{FigLM3}\emph{C}, reveals that heat flow escalates with thermal conductivity, maintaining a tight correlation between theoretical and simulation data. However, the disparity in heat flow between forward and backward directions ($q_{\rm FI}$ and $q_{\rm BIII}$ for inflow, $q_{\rm FIII}$ and $q_{\rm BI}$ for outflow) and its variation with thermal conductivity is less apparent. To address this, we employ Eq.~\textbf{\ref{LM3}} to compute the rectification ratio (Fig. \ref{FigLM3}\emph{D}). We observe a decline in both inflow and outflow rectification ratios with increasing thermal conductivity, where copper demonstrates the lowest ratio and stainless steel the highest.
	
	Further, we produce eight experimental samples, creating two for each type of material. We place two samples simultaneously in opposite directions within the cold and heat sources for each material. Overhead imaging facilitates the documentation of the base plate's surface temperature distribution, presenting results in sequential order: stainless steel, brass, aluminum alloy, and copper (Fig. \ref{FigLM3}\emph{E}).
	Completely isolating the base plate from natural convection during the experiment is challenging. If complete isolation is not achieved and only reduction is possible, the convective coefficient on the base plate is difficult to calculate and control. In practical engineering applications, constructing a vacuum chamber can block the base plate’s natural convection. In our experimental design, multiple samples are needed for parameter discussions, and building multiple vacuum chambers with our current experimental conditions is impractical. Additionally, the presence or absence of natural convection on the base plate does not affect our conclusions on thermal nonreciprocity (\emph{SI Appendix}, Section 10). The main difference lies in whether the heat flow distribution on the base plate remains linear. Thus, we choose to expose the entire structure to air.
	Simulations replicate the experimental conditions, applying identical temperatures for the hot and cold sources. The entire structure is subjected to natural convection in these simulations, adopting a convection coefficient of $h=8~\rm W~m^{-2}~K^{-1}$—a value derived from comparing simulation outputs with experimental observations. The ambient temperature is matched to the experimental setting, recorded at $T_{\rm amb}=301.4$ K.
	
	We align the temperature distribution along the base plate's centerline with the results from simulations (Fig. \ref{FigLM3}\emph{F}). A strong correspondence is observed between the experimental and simulated data. Significant temperature fluctuations in the experimental data can be attributed to temperatures captured at the vertical plates' tops in the overhead images.
	nonreciprocity is markedly evident in the first panel of Fig. \ref{FigLM3}\emph{F}, where the inflow and outflow rectification ratios stand at 0.264 and 0.245, respectively. Yet, with increasing thermal conductivity, the distinction between the forward and backward temperature profiles diminishes progressively. For instance, in the fourth panel of Fig. \ref{FigLM3}\emph{F}, the difference in temperature profiles for the copper sample is scarcely distinguishable, with inflow and outflow rectification ratios recorded at 0.098 and 0.027, respectively. These results, derived from both experiments and simulations, reinforce the principle that nonreciprocal heat transport's intensity inversely correlates with the material's thermal conductivity. While our experiments required substituting the entire structure to vary thermal conductivity, practical engineering applications may achieve nonreciprocal heat transport with adjustable features by merely replacing the vertical plates, leaving the base plate intact.
	
	\begin{figure}[t]
		\includegraphics[width=.9\linewidth]{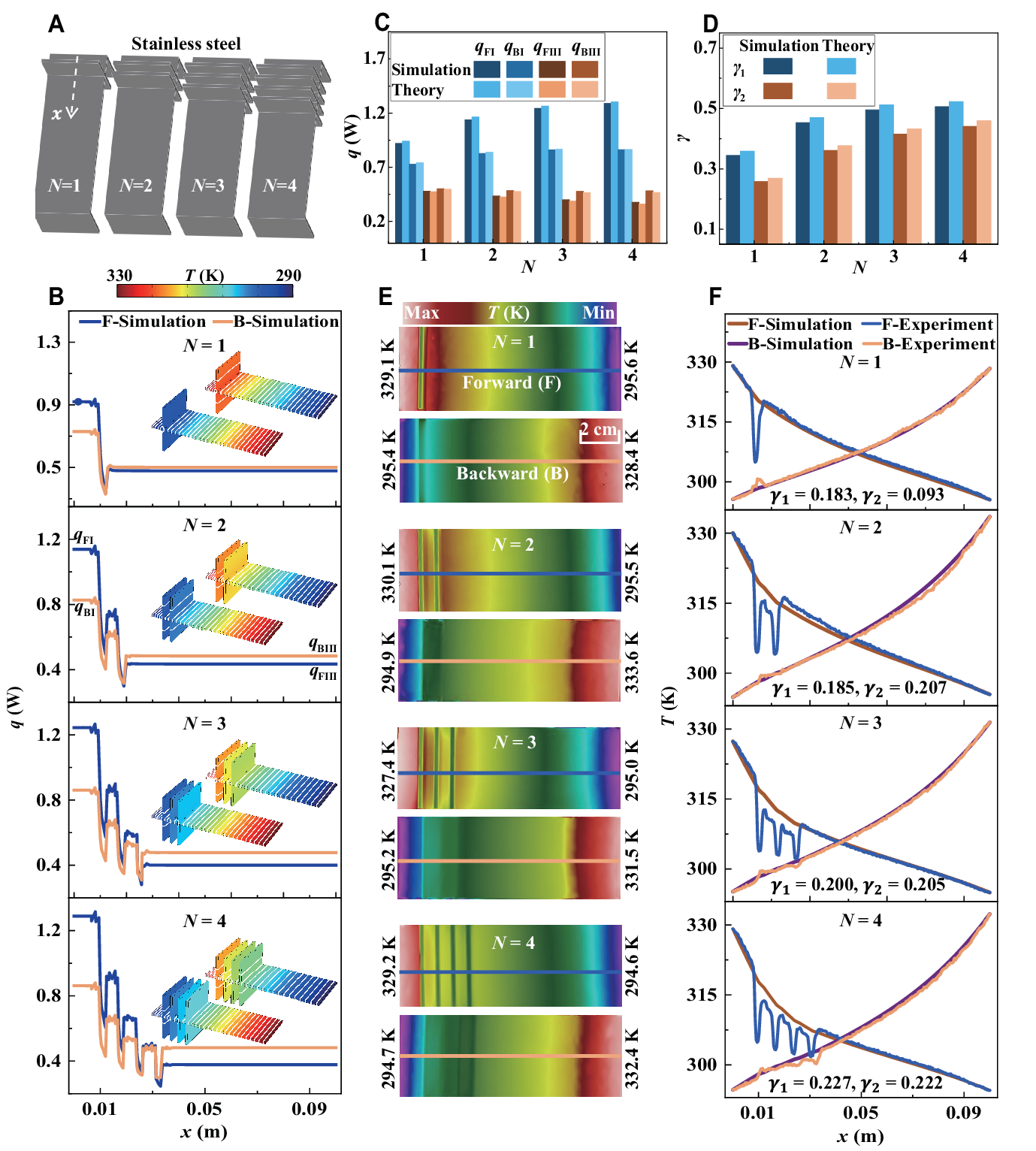}\\
		\caption{Theoretical, simulated, and experimental results of nonreciprocal heat transport configurations with different numbers of vertical plates. ({\it A}) Structural layout. ({\it B}) Simulation outcomes for configurations with 1, 2, 3, and 4 vertical plates, respectively. ({\it C}) and ({\it D}) Heat flow values and the rectification ratio variation with different numbers $N$ of vertical plates. ({\it E}) Experimental temperature distributions for each configuration with 1, 2, 3, and 4 plates, respectively. ({\it F}) Comparative analysis of the temperature distributions along the centerline of the base plate between the experimental data and simulation results. The parameters used in the simulation results are consistent with the experimental parameters.}
		\label{FigLM4}
	\end{figure}
	
	Reconfigurable nonreciprocal heat transport can be realized by altering the number of vertical plates placed on the base plate. In experiments, the stainless steel sample, with its thermal conductivity of $\kappa=16.3~\rm W~m^{-1}~K^{-1}$, demonstrates the highest rectification ratio, showcasing the most distinct nonreciprocal effect. Therefore, we utilize this material to explore how the number of vertical plates affects the rectification ratio.
	Fig. \ref{FigLM4}\emph{A} illustrates a schematic of stainless steel samples featuring various counts of vertical plates. Fig. \ref{FigLM4}\emph{B} shows simulation outcomes for different numbers of plates ($N$), maintaining consistency with the parameters set in Fig. \ref{FigLM3}\emph{B}. These findings reveal that a higher count of plates enhances the discrepancy in heat flow values. By examining the forward and backward heat flow values against theoretical forecasts and subsequently computing the rectification ratio (Fig. \ref{FigLM4} \emph{C}-\emph{D}), we observe that both the inflow and outflow rectification ratios directly correlate with the plate count. Notably, the rectification ratio undergoes significant shifts up to three plates, with minimal variations observed from three to four plates. This negligible change is linked to the diminished difference in natural convection during forward and backward heat transport by the fourth plate, located nearer to the center, hence its lesser impact on the rectification ratio.
	
	Fig. \ref{FigLM4}\emph{E} presents experimental data under natural convection conditions for the entire structure. Fig. \ref{FigLM4}\emph{F} compares the simulation and experimental temperature profiles along the structure's centerline, applying uniform parameters including a natural convection coefficient of $h=8~\rm W~m^{-2}~K^{-1}$ and an ambient temperature of $T_{\rm amb}=301.4$ K. The temperature distribution disparity escalates with the plate count, closely mirroring the simulation and experimental findings (Fig. \ref{FigLM4}\emph{F}). According to simulations, the inflow rectification ratio rises with the plate count, whereas the outflow rectification ratio experiences minor variations. These variations result from experimental conditions not being strictly controlled, leading to slight inconsistencies in the temperatures of the hot and cold sources across trials. With rigorous variable control (\emph{SI Appendix}, Section 10), the inflow and outflow rectification ratios for the stainless steel samples proportionally increase with the plate count.
	This investigation affirms that by adjusting the thermal conductivity and plate count, reconfigurable nonreciprocal heat transport is attainable. Additional factors, such as ambient temperature and the temperatures of the heat and cold sources, further influence nonreciprocal heat transport. This approach enables reconfigurable nonreciprocal heat transport without necessitating structural modifications, ensuring the outcomes' universality and robustness.

	\section*{Conclusion and discussion}
	
	We've successfully demonstrated reconfigurable, zero-energy, and wide-temperature nonreciprocal heat transport using asymmetric structures made of natural bulk materials, capitalizing on heat loss differentials. Asymmetric natural convection heat loss between vertical plates and surrounding air creates spatial heat conduction asymmetry. A precise mathematical model theoretically connects the rectification ratio to various parameters. A rectification ratio greater than 0.1 can be considered effective thermal nonreciprocity. Due to the limitations of the structure’s thermal conductivity, our adjustable rectification ratio range is limited, making it difficult to reach the maximum value of 1. However, the optimal rectification ratio can be achieved through the combined adjustment of multiple parameters. The rectification ratio increases with the thermal conductivity of the vertical plates, decreases with the overall structure’s thermal conductivity, and increases with the number of vertical plates. 
	Simulations and experimental data show a reconfigurable rectification ratio consistent with our theoretical predictions. Reconfigurable thermal nonreciprocal metamaterials can guide the design of multifunctional metamaterials, such as thermoelectric materials \cite{MaoNM21}, which optimize energy collection and conversion by managing heat flow direction and intensity.
	
	Thermal nonreciprocity is driven solely by ambient temperature differences, allowing the rectification ratio to be tailored to environmental changes. For example, as shown in Fig. \ref{FigLM2}\emph{H}, with a temperature difference of 100 K between the cold and hot sources, the nonreciprocal effect is significant within an ambient temperature range of 286 K to 360 K. Similarly, at a typical ambient temperature of 298 K, nonreciprocity is maintained across various temperature differences between the sources, as illustrated in Fig. \ref{FigLM2}\emph{I}. By jointly adjusting the ambient temperature and the temperature difference between the cold and hot sources, zero-energy nonreciprocity can be achieved over a wide temperature range.
	Zero-energy nonreciprocal designs can guide the creation of energy-saving materials. For example, thermally nonreciprocal metamaterials enabled by extended wall design can replace high-energy-consuming equipment like air conditioners to provide energy-free temperature control in enclosed spaces (\emph{SI Appendix}, Section 11). Taking a greenhouse as an example, the combination of high-transmittance materials and an enclosed structure allows the indoor temperature to exceed that of the outdoor environment. Seasonal temperature differences lead to varying heat accumulation indoors. Excessive heat in summer and insufficient warmth in winter can inhibit plant growth, creating different thermal management needs across seasons. Our adjustable nonreciprocal structure design can help dissipate heat in summer and retain it in winter, enabling effective temperature regulation throughout the year. This approach achieves efficient temperature control without consuming external electric energy, significantly enhancing building energy efficiency \cite{AbuAE19}. 
	The directional heat transport characteristics of thermal nonreciprocity can also guide hyperthermia. In hyperthermia \cite{ZhuCR23}, the human body remains comfortable around 310 K, and temperatures above 312 K negatively impact cancer cells. An asymmetric structure made of bionic material is implanted in the human body, and the convection heat transfer in the blood or the conduction heat transfer between tissues is used to selectively heat cancer cells under a high-temperature heat source to inhibit their growth (\emph{SI Appendix}, Section 11). The wide temperature range design of thermal nonreciprocity can adjust rectification ratios based on the type of injury and diseased cells, ensuring maximum comfort and treatment efficacy. Currently, the fabrication and realization of thermal nonreciprocity remain at the experimental stage, distant from real applications. Discussing practical applications of nonreciprocal design in buildings and medicine can help bridge this gap.
	
	\section*{Materials and Methods}
	
	\subsection*{Finite-Element Simulations}
	We utilize Comsol Multiphysics software, modeling heat transport with solid heat transport plates. We simulate natural convection by applying convective heat flux across the structure's surfaces. In the configurations depicted in Fig. \ref{FigLM2} \emph{B}-\emph{I}, \ref{FigLM3}\emph{B}, \ref{FigLM4}\emph{B}, the surfaces at $x=0$ and $x=c$ serve as the cold and hot source surfaces, respectively. All surfaces of the vertical plates are subject to convective heat flux while the remaining surfaces are treated with thermal insulation. For the designs showcased in Fig. \ref{FigLM3}\emph{F} and \ref{FigLM4}\emph{F}, convective heat flux is implemented on all surfaces except those at $x=0$ and $x=c$.
	
	The structures in Fig. \ref{FigLM3} and \ref{FigLM4} share the following geometric parameters: a base plate length of 0.1 m, a width of 0.04 m, and a thickness of 0.002 m. For the Fig. \ref{FigLM3} structure, the combined length of the vertical plates is 0.042 m, symmetrically aligned along the base plate. The $x$-axis positions of the five vertical plates are 0.011 m, 0.018 m, 0.025 m, 0.032 m, and 0.039 m, each with a thickness of 0.002 m. The structure in Fig. \ref{FigLM4} maintains identical geometric specifications as Fig. \ref{FigLM3}, differing only in the reduced count of vertical plates starting from the center.
	
	\subsection*{Experimental Demonstration}
	The experimental design replicates the numerical simulation's shape and boundary conditions, as shown in Fig. \ref{FigLM3}\emph{A} and \ref{FigLM4}\emph{A}, with the setup detailed in \emph{SI Appendix}, Section 12. To capture high-quality thermal images, we apply transparent book covers to all sample surfaces to reduce the reflectivity of the metal surfaces. Compared to the simulated structure, the experimental model extends both ends by 0.03 m and features bends. These extensions are submerged in cold and hot water baths to mimic the simulation's hot and cold source temperatures, maintained with a temperature-controlled heating rod and an ice-water mixture, respectively.
	
	To ensure accurate temperature replication for both forward and backward heat transport, we use two identical samples for each experimental parameter, arranging the cold and heat sources in opposite directions. The experiments are performed in an air-conditioned space to ensure a stable ambient temperature, thus reducing variations in the natural convection coefficient. By positioning the experiments away from air vents, we further guarantee a uniform natural convection setting.
	The convective coefficient in the experiments is derived from fitting simulation and experimental results. Considering the stainless steel sample in Fig. 3A, we obtain the temperature distribution and the center temperature profile along the $x$-axis. All size and temperature parameters used in the experiment are input into the simulation. In the experiment, we can not completely eliminate the impact of thermal radiation from the metal surface. This thermal radiation \cite{ZhangCPL23}, also considered a thermal loss, helps achieve thermal nonreciprocity. Thus, we equate all heat exchange between the air and the vertical plate to convective heat flux applied to the sample surface in the simulation. In the simulation, all parameters are known except for the convective coefficient. The typical magnitude of the air’s natural convective coefficient is 5 to 25 $\rm W~m^{-2}~K^{-1}$. We vary the heat transfer coefficient until the simulated temperature curve matches the experimental curve, at which point the heat transfer coefficient is determined to be our equivalent convective coefficient ($h=8~\rm W~m^{-2}~K^{-1}$). All experiments are conducted in the same environment with consistent structural parameters for the vertical plates, ensuring that the equivalent convective coefficient is the same for all samples. The perfect agreement between simulation and experimental results validates this approach. 
	
	\subsection*{Data Availability}
	All study data are included in the article and/or SI Appendix.

	\section*{Acknowledgments}
	J. H. gratefully acknowledged the support by the National Natural Science Foundation of China through Grants No. 12035004 and No. 12320101004, and the Innovation Program of the Shanghai Municipal Education Commission with Grant No. 2023ZKZD06. L. X. acknowledged the support by the National Natural Science Foundation of China under Grants No. 12375040, No. 12088101, and No. U2330401. Y. L. acknowledged the support by the National Natural Science Foundation of China under Grants No. 92163123 and No. 52250191, and Zhejiang Provincial Nat
	ural Science Foundation of China under Grant No. LZ24A050002.

\end{document}